\documentclass[runningheads]{llncs}
\usepackage{graphicx}
\usepackage[linesnumbered,ruled,vlined]{algorithm2e}
\usepackage{booktabs} % For formal tables
\usepackage{amsfonts}
\usepackage{cite}
 
\newcommand{\squishlist}{
    \begin{list}{$\bullet$}
    { \setlength{\itemsep}{0pt}
        \setlength{\parsep}{1pt}
        \setlength{\topsep}{1pt}
        \setlength{\partopsep}{0pt}
        \setlength{\leftmargin}{1em} %1.5em
        \setlength{\labelwidth}{1em}
        \setlength{\labelsep}{0.5em}
    						 } }

\newcommand{\squishlisttwo}{
    \begin{list}{$\bullet$}
        { \setlength{\itemsep}{0pt}
            \setlength{\parsep}{0pt}
            \setlength{\topsep}{0pt}
            \setlength{\partopsep}{0pt}
            \setlength{\leftmargin}{2em}
            \setlength{\labelwidth}{1.5em}
            \setlength{\labelsep}{0.5em} } }

\newcommand{\squishend}{
    \end{list}  }
\begin{document}
\title{Evolution of Filter Bubbles and Polarization in News Recommendation}

\author{Han Zhang\inst{1}\orcidID{0000-0002-9695-5658}\and
Ziwei Zhu\inst{2}\orcidID{0000-0002-3990-4774}\and
James Caverlee\inst{1}\orcidID{0000-0001-8350-8528}}

% % \authorrunning{F. Author et al.}

\institute{Texas A\&M University, 400 Bizzell St College Station, USA \and
George Mason University, 4400 University Drive Fairfax, USA}

\maketitle              % typeset the header of the contribution
\begin{abstract}

Recent work in news recommendation has demonstrated that recommenders can over-expose users to articles that support their pre-existing opinions. However, most existing work focuses on a static setting or over a short-time window, leaving open questions about the long-term and dynamic impacts of news recommendations. In this paper, we explore these dynamic impacts through a systematic study of three research questions: 1) How do the news reading behaviors of users change after repeated long-term interactions with recommenders? 2) How do the inherent preferences of users change over time in such a dynamic recommender system? 3) Can the existing SOTA static method alleviate the problem in the dynamic environment? Concretely, we conduct a comprehensive data-driven study through simulation experiments of political polarization in news recommendations based on 40,000 annotated news articles. We find that users are rapidly exposed to more extreme content as the recommender evolves. We also find that a calibration-based intervention can slow down this polarization, but leaves open significant opportunities for future improvements

\keywords{Filter bubble  \and Recommender system \and Dynamic.}
\end{abstract}
\section{Introduction}
\label{sec:intro}

% The past decade has witnessed the thriving development of personalized recommender systems, which offer the promise of improved user experience. However, 

% existing work~\cite{nguyen2014exploring, cinelli2021echo} that personalized recommender systems can raise issues of echo chambers and filter bubbles. Indeed,

% For example, how fast do these filter bubbles form? Does polarization oscillate? Or is it fixed? And can interventions alleviate this polarization? 
% And if so, how long do they take to show impact? 

It has been demonstrated by recent work\cite{pariser2011filter,liu2021interaction} that \textit{personalized news recommender systems} can over-expose users to news articles supporting their pre-existing opinions. With increasing reliance on personalized recommendations to consume news from digital news apps\cite{agarwal2009explore,das2007google}, such a filter bubble phenomenon paves the way for continued (and potentially increased) intellectual segregation and political polarization.

While these important studies have demonstrated the problem of filter bubbles and political polarization, most existing work\cite{rodriguez2017partisan,abdollahpouri2019unfairness,park2008long,steck2011item} focuses on the problem under a static or short-term setting, leaving open questions about the \textit{long-term and dynamic impacts} of news recommendations.  For example, how fast do these filter bubbles form? Does polarization oscillate? Or is it fixed? Can interventions alleviate this polarization? Hence, in this work, we conduct a systematic study to investigate the long-term and dynamic impacts of news recommender systems organized around three key research questions: 1) How do the news reading behaviors of users change after repeated long-term interactions with recommenders? 2) How do the inherent preferences of users change over time in such a dynamic recommender system? 3) Can a SOTA intervention method alleviate the problem in the dynamic environment?

Concretely, we conduct an extensive data-driven study through simulations of news recommendations based on 40,000 annotated news articles to study the impacts of news recommenders. To uncover how the recommender influences the news reading behaviors of users and intensifies polarization over time, we consider that the political preferences of users can be influenced by recommended and read news. Unsurprisingly, we find that users are rapidly exposed to more extreme content as the recommender evolves and the inherent political preferences of users become increasingly radical. Moreover, we also observe that users read more and more extreme content even if they are immune to recommendation influence and keep their inherent political opinions invariant. Last, we further conduct experiments with a calibration-based method~\cite{steck2018calibrated}, which is a SOTA static method for addressing filter bubbles. We find that such a calibration-based intervention can slow down this polarization but still leaves open significant opportunities for future improvements.

\section{Dynamic Experiment Setup}
\label{sec:Experiment Setups}

In this section, we first introduce our framework for studying dynamic news recommendation, including the dataset, the experimental process, and the metrics. 

% Then, we show how recommendations influence news reading behaviors of users and lead to political polarization.% by dynamic experiments.

\subsection{Dataset}

\begin{figure}[t!]
% \vspace{-12pt} 
\centering
\includegraphics[ width=1\linewidth ]{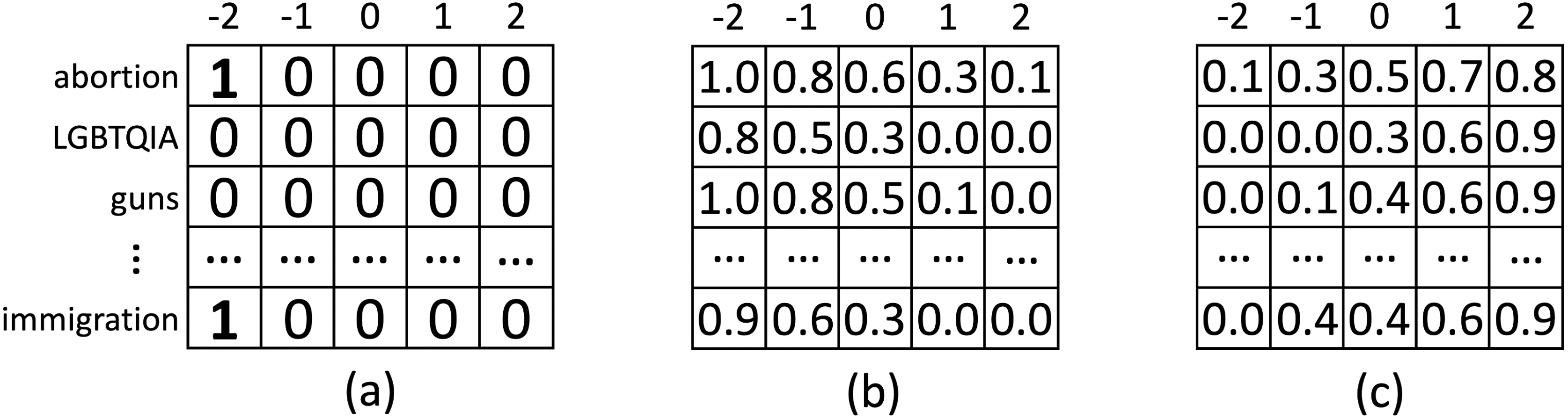} 
% \vspace{-10pt} 
\caption{(a) shows an article matrix. (b) shows a preference matrix for a `solid liberal' user. (c) shows a preference matrix for a `core conservative' user.}
\label{fig:matrix-examples} 
% \vspace{-15pt} 
\end{figure}

% Fig.~\ref{fig:matrix-examples} shows the relationship between political stance values and political group names.

We use a variation of the dataset from \cite{liu2021interaction}, which consists of a collection of 40,000 news articles and a set of 500 users. The 40,000 articles are with annotations of their topics and political stances. Specifically, there are 14 topics: \textit{abortion, environment, guns, health care, immigration, LGBTQIA, taxes, technology, trade, Trump impeachment, US military, welfare, US 2020 election, and racism}. Each article can cover one or more topics. For political stance, each article is labeled as one of $\{-2, -1, 0, 1, 2\}$, which spans the ideological spectrum from extreme liberal $(-2)$ to extreme conservative $(+2)$. There are 8,000 articles for each political stance. We can use a binary utility matrix $\mathbf{A}_i\in\{0,1\}^{14\times5}$ to represent the topic and stance for an article $i$. Figure~\ref{fig:matrix-examples}(a) shows an example of an article related to abortion and immigration with a political stance of -2.

The user set is simulated based on the Pew survey of U.S. political typologies~\cite{doherty2017political}, which summarizes 9 political typologies in the U.S. and their opinions toward different topics. We consider the five most representative typologies: \textit{solid liberal} (extreme liberal), \textit{opportunity democrats} (lean toward liberal), \textit{bystanders} (mild group), \textit{market skeptic republicans} (lean toward conservative), and \textit{core conservatives} (extreme conservative). For each typology, we generate 100 users, where each user can be represented by a preference matrix $\mathbf{U}_u\in\mathbb{R}^{14\times5}$ to represent the  user's political stances toward different topics. The larger $\mathbf{U}_u(t,s)$ is, the more likely user $u$ holds an opinion of stance $s$ toward the topic $t$. Figure~\ref{fig:matrix-examples}(b) shows an example preference matrix of a `solid liberal' user and Figure~\ref{fig:matrix-examples}(c) shows an example preference matrix for a `core conservative' user. 

% The larger the value is, the more likely the user holds political stance $s$ toward topic $t$.

With the utility matrices for news articles and preference matrices of users, we can determine the preference of a user for an article by vectorizing their corresponding matrices and then taking the dot product to calculate the preference score between them. We can further use this preference score to determine user-read-article behaviors. The higher the preference score is, the more likely a user is to click and read the article. More details about how news articles are annotated, how user preference matrices are generated from the Pew survey, and the user click model can be found in \cite{liu2021interaction}.

\setlength{\textfloatsep}{6pt}
\subsection{Dynamic Recommendation Process}
\label{sec:Long-term Experiment Process}
\setlength{\textfloatsep}{6pt}

% \begin{wrapfigure}{R}{0.62\textwidth}
%     \begin{minipage}{0.62\textwidth}
%         \vspace{-15pt}    
%         \begin{algorithm}[H]
%         \textbf{Bootstrap:} Randomly expose 10 articles from each topic (140 in total) to each user, and collect initial clicks $\mathcal{D}$, and train the first model $\psi$ by $\mathcal{D}$\;
%         \For{$t=1:40,000$}{
%             Randomly choose a user $u_t$ as the current visiting user\;
%             Recommend 5 articles to the current user $u_t$ by $\psi$\;
%             Collect new clicks and add them to $\mathcal{D}$\;
%             Update preference matrix of user $u_t$\;
%             \If{$t\%200==0$}{
%                 Retrain $\psi$ by $\mathcal{D}$\;
%             }
%         }
%         \caption{Dynamic News Recommendation}
%         \label{alg:dynamic_rec} 
        
%         \end{algorithm}
% \end{minipage}
% \end{wrapfigure}

\begin{algorithm}[t!]
\textbf{Bootstrap:} Randomly expose 10 articles from each topic (140 in total) to each user, and collect initial clicks $\mathcal{D}$, and train the first model $\psi$ by $\mathcal{D}$\;
\For{$t=1:40,000$}{
    Randomly choose a user $u_t$ as the current visiting user\;
    Recommend 5 articles to the current user $u_t$ by $\psi$\;
    Collect new clicks and add them to $\mathcal{D}$\;
    Update preference matrix of user $u_t$\;
    \If{$t\%200==0$}{
        Retrain $\psi$ by $\mathcal{D}$\;
    }
}
\caption{Dynamic News Recommendation}
\label{alg:dynamic_rec} 
\end{algorithm}
\setlength{\textfloatsep}{6pt}

Next, we conduct a dynamic recommendation experiment to study how users are impacted by a personalized news recommender. The detailed experimental process is presented in Algorithm~\ref{alg:dynamic_rec}. We first conduct a bootstrap step to collect initial click data from all users by randomly showing 140 articles (10 articles from each topic) and then training the first recommendation model with the initial click data. Then, we run the dynamic experiment for 40,000 iterations. At each iteration, a random user will come and ask for recommendations of 5 articles. The user will iterate all the 5 articles and determine whether click and read them. The interaction data will be stored for further model training. We retrain the model after every 200 iterations, resulting in 200 experiment epochs. In this work, we use the fundamental Matrix Factorization (MF)~\cite{liu2021interaction} model as the core approach to deliver recommendations.

Moreover, in the real world, users' preferences can be influenced by recommendations exposed to them. If an article was recommended and read by a user, the corresponding opinions of the user will be reinforced, and the user is more likely to click articles with the same political stances and topics in the future. So, we model these dynamics by changing preference matrices of users corresponding to what articles are exposed and read by users. We first define an influence parameter $c$ to determine to what degree users can be influenced by recommendations. Then, every time a user $u$ is exposed to an article $i$, if $u$ clicks and reads $i$, we update the preference matrix $\mathbf{U}_u$ of $u$ by $\mathbf{U}_u\leftarrow \mathbf{U}_u + c\cdot \mathbf{A}_i$. A larger $c$ means that people are more susceptible to the recommendation influence.

\begin{figure*}[t!]
\vspace{0pt}
\centering
\includegraphics[ width=1\linewidth ]{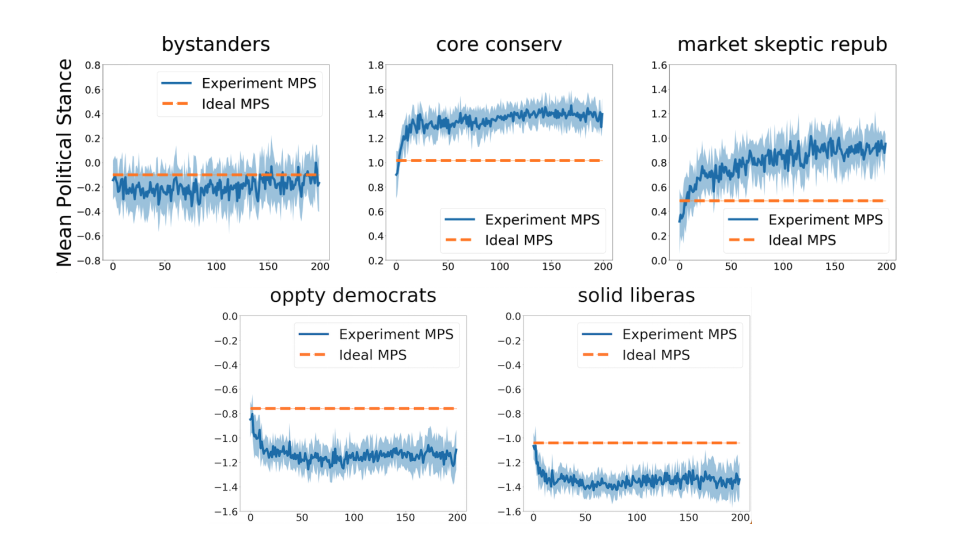} 
% \vspace{-20pt} 
\caption{MPS changes over time for five user groups (c = 0). }
\label{fig:MPS_0} 
% \vspace{-5pt} 
\end{figure*}

\subsection{Evaluation Metrics}
\label{sec:metrics}

% In this work, we use the Discounted Cumulative Gain (DCG)~\cite{busa2012apple, dupret2011discounted} to evaluate the utility of the recommender system: $DCG_t = \sum_{p = 1}^{5}\frac{y_{u_t,p}}{log_2(p+1)}$. After recommending 5 articles to a user, we iterate the 5 recommended articles (from top position $p=1$ to the end $p=5$), and if $u_t$ clicks and reads article at position $p$, $y_{u_t,p}=1$, otherwise $y_{u_t,p}=0$. In the result, we report the average DCG for one experiment interaction, which contains 200 iterations. 
% The change of average DCG over the 40,000 iterations will show how the recommender system works on different user group.

To show how recommendations influence user behavior, we calculate the Mean Political Stance (MPS) for iteration $t$: $MPS_t = \sum_{p = 1}^{5}y_{u_t,p}\cdot {stance(p)}/{\sum_{p = 1}^{5}y_{u_t,p}}$, where we iterate the 5 recommended articles (from top position $p=1$ to the end $p=5$), and if $u_t$ clicks and reads article at position $p$, $y_{u_t,p}=1$, otherwise $y_{u_t,p}=0$. We calculate the average political stance of articles read by the user at interaction $t$, and $stance(p)$ returns the political stance of an article at position $p$. We report the average MPS for each user group in each experiment epoch and show how it evolves over 200 epochs.

We also calculate the User Mean Political Stance (UMPS) for each user: $UMPS = \sum_{s = -2}^{2} {\sum_{t = 1}^{14}s\cdot \mathbf{U}_u(t,s)}$ to directly show the evolution of inherent user preference. The UMPS reflects the current user preference. After each epoch, we calculate the average UMPS of each user group and show how their inherent preference change over 200 epochs.

\section{Experimental Results}
We empirically study three key research questions: (RQ1) In such a dynamic recommendation process, will users be exposed to and read more and more similar articles with more extreme political stances? (RQ2) Will users be influenced by these recommendations and become more and more radical over time? and (RQ3) Can an existing intervention method alleviate the filter bubble problem? All experiments are repeated 10 times, and we report the averaged results. 
% \textit{All code and data can be found at http://anonymous.url}.

\begin{figure*}[t!]
% \vspace{10pt}
\centering
\includegraphics[ width=1.\linewidth ]{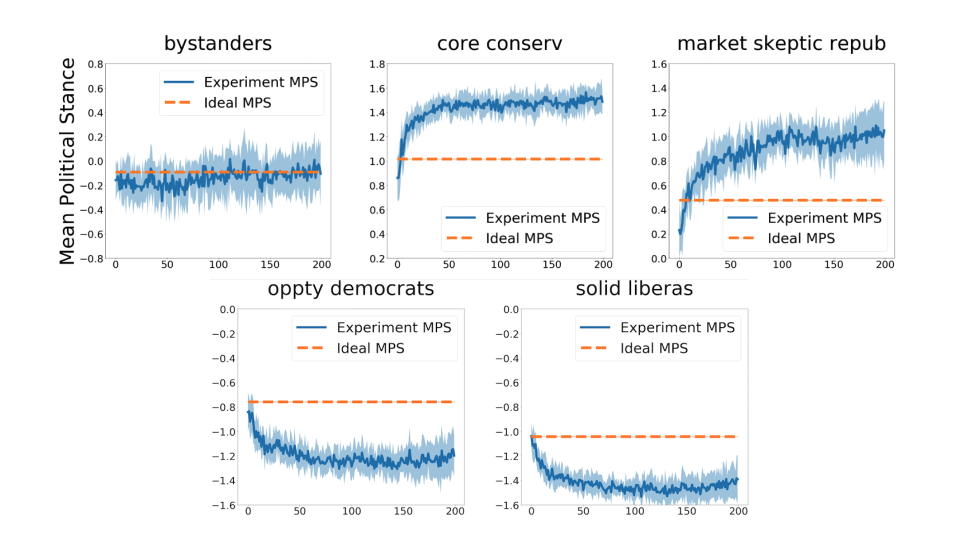} 
% \vspace{-20pt} 
\caption{MPS changes over time for five user groups (c = 0.03). }
\label{fig:MPS_0.03} 
% \vspace{-5pt} 
\end{figure*}

% \medskip
\smallskip
\noindent \textit{\textbf{RQ1: Evolution of User News Reading Behaviors.}} First, we study how do the news reading behaviors of users change after repeated long-term interactions with the recommender. We report the averaged MPS of each user group to depict the pattern of news reading behaviors, and show the changing of behavior patterns with influence parameter $c=0$ in Fig.~\ref{fig:MPS_0} and with $c=0.03$ in Fig.~\ref{fig:MPS_0.03}. The x-axis in these figures represents the experiment epochs, each of which contains 200 interactions. In the figures, besides the MPS in each epoch during the experiment, we also plot the MPS during the bootstrap step for each user group, which indicates the true initial political stance of each user group and can be regarded as the ideal MPS we want to achieve for dynamic recommendation.

From the result, we can see that even though user political preference remains static (the influence parameter is set to be 0), the absolute value of MPS of different groups except the `bystanders' group becomes larger. In other words, even if users are immune to the influence of recommendations and keep their political preferences invariant, they will still read more and more extreme news. After we add the influence parameter into the experiment and compare the results in Figure~\ref{fig:MPS_0} and Figure~\ref{fig:MPS_0.03}, we can observe even more severe trend of reading extreme news: except for the `bystanders' group, the other four groups become more and more deviated from the ideal MPS, demonstrating the rapid trend of radicalization and polarization of users.

\begin{figure}[t]
\centering
% \vspace{-5pt}
\includegraphics[ width=0.9\linewidth ]{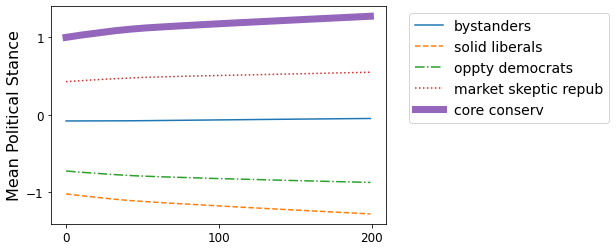} 
% \vspace{-10pt} 
\caption{The UMPS of different user groups change over time (c=0.03).}
\label{fig:UMPS} 
% \vspace{-2pt} 
\end{figure}

\smallskip
\noindent \textit{\textbf{RQ2: Evolution of User Preference.}} Next, we unveil how the inherent political preferences of users evolve over time. Here, we measure the averaged UMPS for each user group to indicate the current inherent preference of the user group and show the changing of UMPS over time to depict how the user preference is influenced by recommendations. In Figure~\ref{fig:UMPS}, we show the empirical result with $c=0.03$, which clearly illustrates that except for `bystanders', all other user groups become more and more radical. That is to say, if the exposed recommendations can change users' inherent opinions, users will move toward more extreme stances after long-term interactions with the new recommender.%, raising the political polarization issue substantially.

\begin{figure}[t]
\centering
% \vspace{-5pt}
\includegraphics[ width=0.9\linewidth ]{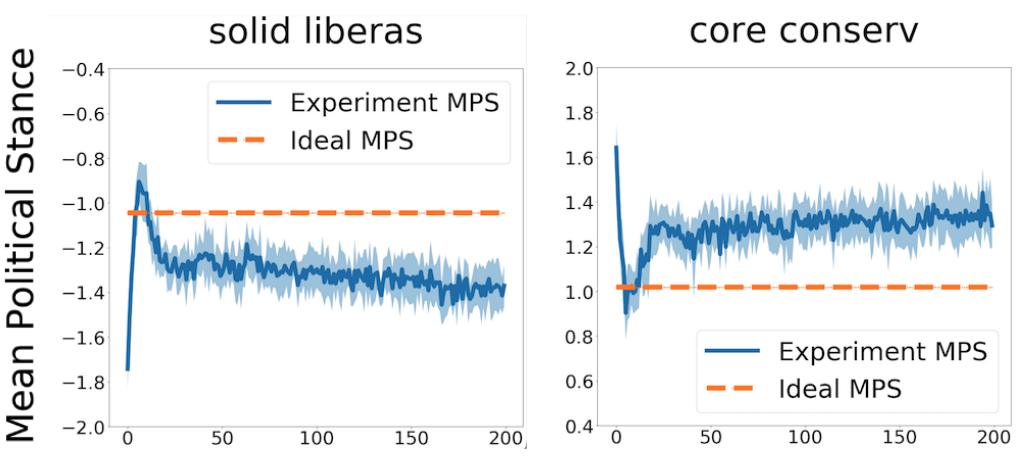} 
% \vspace{-10pt} 
\caption{Changing of MPS with the Calibrated Recommendation method (c=0.03)}
\label{fig:calibrated} 
% \vspace{2pt} 
\end{figure}

\smallskip
\noindent \textit{\textbf{RQ3: Effectiveness of Intervention.}} Last, we study how well a SOTA static method for mitigating filter bubble performs for the dynamic recommendation. Here, we conduct experiments to evaluate the performance of the Calibrated Recommendation method~\cite{steck2018calibrated}, which is one of the SOTA static methods for addressing filter bubbles. The Calibrated method re-ranks the recommendation list from the recommender so that the re-ranked list contains a distribution that follows an ideal distribution (the distribution learned from the bootstrap step in our case). We show the results for the ``solid liberals" and ``core conservatives" groups with $c=0.03$ in Fig.~\ref{fig:calibrated}. From these results, we can observe that the Calibrated method can only slow down the polarization process, but it cannot prevent the trend of radicalization and polarization. Hence, we conclude that the calibration method produces very limited effects in such a dynamic scenario motivating efforts for more effective methods.

% to the dynamic filter bubble and polarization problem, and a more effective method that can dynamically alleviate the problem is in urgent demand.

\section{Related Work}
Filter bubble is a long-standing problem for recommender systems, widely studied in many large-scale platforms like Twitter, Facebook, and YouTube \cite{barbera2015tweeting,bakshy2015exposure,eady2019many,min2018all,shapiro2015more}. One of the major reasons raising filter bubbles is the nature of recommendation algorithms to deliver content that users are more likely to click on to maximize utility\cite{chu2009personalized,dumais2003sigir,johnson2014logistic}. Such a problem of filter bubbles can lead to damaged user experience and intensify intellectual segregation and polarization in society~\cite{epstein2015search}. Specifically, a recent work~\cite{liu2021interaction} analyzes and compares how different algorithms form filter bubbles and expose more extreme content to users in a news recommender system. However, most prior work is focused on short-term and static scenarios, which motivates us to explore the long-term and dynamic nature of filter bubbles in this work.

\section{Conclusion and Future Work}
In this paper, we conduct a comprehensive data-driven study through simulation experiments of political polarization in news recommendations based on 40,000 annotated news articles. Specifically, we answer three research questions: 1) How do the news reading behaviors of users change after repeated long-term interactions with recommenders? 2) How do the inherent preferences of users change over time in such a dynamic recommender system? 3) How effective can the existing SOTA intervention method alleviate the problem in the dynamic environment? We find that users are rapidly exposed to more extreme content and become more radical as the system evolves. We also find that a calibration-based intervention slows down this polarization, but leaves open significant opportunities for future improvements

\section{Acknowledgements}
This work is supported in part by NSF grants IIS-1939716 and IIS-1909252.

\newpage
\clearpage
\bibliographystyle{splncs04}
\bibliography{samplebib}
\end{document}